# SWITCH: An Exemplar for Evaluating Self-Adaptive ML-Enabled Systems

Arya Marda
arya.marda@students.iiit.ac.in
Software Engineering Research
Center, IIIT Hyderabad
India

Shubham Kulkarni
shubham.kulkarni@research.iiit.ac.in
Software Engineering Research
Center, IIIT Hyderabad
India

Karthik Vaidhyanathan
karthik.vaidhyanathan@iiit.ac.in
Software Engineering Research
Center, IIIT Hyderabad
India

## ABSTRACT

Addressing runtime uncertainties in Machine Learning-Enabled Systems (MLS) is crucial for maintaining Quality of Service (QoS). The Machine Learning Model Balancer is a concept that addresses these uncertainties by facilitating dynamic ML model switching, showing promise in improving QoS in MLS. Leveraging this concept, this paper introduces SWITCH, an exemplar developed to enhance self-adaptive capabilities in such systems through dynamic model switching in runtime. SWITCH is designed as a comprehensive web service catering to a broad range of ML scenarios, with its implementation demonstrated through an object detection use case. SWITCH provides researchers with a flexible platform to apply and evaluate their ML model switching strategies, aiming to enhance QoS in MLS. SWITCH features advanced input handling, real-time data processing, and logging for adaptation metrics supplemented with an interactive real-time dashboard for enhancing system observability. This paper details SWITCH's architecture, self-adaptation strategies through ML model switching, and its empirical validation through a case study, illustrating its potential to improve QoS in MLS. By enabling a hands-on approach to explore adaptive behaviors in ML systems, SWITCH contributes a valuable tool to the SEAMS community for research into self-adaptive mechanisms for MLS and their practical applications.

## CCS CONCEPTS

• **Software and its engineering** → **Software system structures**; **Designing software**;

## KEYWORDS

Self-Adaptation, Self Adaptive Systems, Machine Learning-Enabled Systems, Exemplar, Web Service, Object Detection

## 1 INTRODUCTION

Machine Learning-Enabled Systems (MLS), such as Google Bard and ChatGPT, are increasingly prevalent, offering a diverse array of services powered by advancements in AI. However, these systems encounter runtime uncertainties due to environmental factors, changing conditions, etc. They also face additional challenges unique to their data-driven nature, where outcomes and performances are inherently variable and highly dependent on data quality, ML model design and accuracy [2]. This evolving AI landscape in MLS underscores the necessity for adaptive software architecture [20] to effectively manage these uncertainties [8]. Traditionally, research in self-adaptive systems (SAS) has focused on non-ML-based systems, emphasizing tactics to adapt software architecture or configurations in response to environmental uncertainties [8, 24]. However, these approaches often fall short of addressing the complex and evolving requirements of MLS. Recent advancements highlight the growing importance of integrating SAS principles with AI technologies, especially in the context of MLS [1], to effectively manage these new challenges.

In our previous work [15], we introduce the Machine Learning Model Balancer concept, advocating for dynamic switching between ML models during runtime to manage uncertainties and optimize Quality of Service (QoS). In response to these challenges, we present SWITCH, an exemplar developed leveraging the Machine Learning Model Balancer concept, designed to allow researchers and practitioners to explore and refine self-adaptive strategies specifically in the context of MLS. This approach demonstrates how effectively switching between different ML models in runtime in response to operational demands can enhance system performance, recognizing the potential of self-adaptation in MLS. The practicality of this concept, as shown in AdaMLS, highlights a gap in the SEAMS community[1]: the absence of tools for experimenting with and refining self-adaptive strategies in the context of ML's unique, data-driven, and non-deterministic nature. While existing literature and exemplars within the SEAMS community provide insights, they primarily focus on non-ML systems scenarios.

To address this gap in this work, we present SWITCH, an exemplar of real-world ML-enabled systems. It is demonstrated through applications in the object detection domain. SWITCH enables runtime ML model switching and offers an end-to-end platform for handling varying loads, data drifts, and various uncertainties in MLS. Key features include input handling, real-time data processing, and a user-friendly dashboard, facilitating effective monitoring and experimentation in real-world scenarios. SWITCH is used in validating the AdaMLS, self-adaptation approach, illustrating its capacity to enhance QoS in MLS. Thus, SWITCH stands as a valuable tool for the self-adaptation and MLS research community, providing a unique and practical platform where researchers can thoroughly test, analyze, and refine their self-adaptation strategies, ensuring their efficacy and effectiveness before deployment in real-world ML scenarios. SWITCH exemplar is available in Git Repository[2] and official website[3]. A video demonstration of the tool is also made available[4]. The remainder of the paper is structured as follows. Section II provides an overview of SWITCH. Section III delves into the architecture and design of SWITCH. Section IV discusses System Usage and Adaptation. Section V presents an empirical validation case study and focuses on technical challenges and solutions. Section VI discusses related work, and Section VII gives future directions and presents the conclusions.

---

[1]https://www.hpi.uni-potsdam.de/giese/public/selfadapt/exemplars/
[2]https://github.com/sa4s-serc/switch
[3]https://tool-switch.github.io
[4]https://www.youtube.com/@tool-switch

## 2 OVERVIEW

SWITCH is designed as a practical web service for ML, functioning in an online deployment mode. It stands out as an exemplar in the field of MLS, offering a unique simulation platform for dynamic model switching through software architecture-based self-adaptation through MAPE-K framework [23]. This approach effectively addresses the challenge of maintaining Quality of Service (QoS) in the face of operational uncertainties. The system's architecture is tailored for handling complex ML scenarios, particularly demonstrated through object detection use cases. SWITCH integrates input handling via FastAPI[2], facilitating seamless and efficient user interactions.

It employs state-of-the-art YOLOv5u object detection models [12] for real-time data processing, ensuring high accuracy and responsiveness. The system also features systematic logging of observability metrics and system logs in Elasticsearch[2], providing a robust framework for data management & analysis. A standout feature of SWITCH is its interactive, real-time dashboard, implemented using Kibana[2]. This user-friendly interface is designed for effective experiment management and system performance monitoring. It allows researchers to visualize the model switching process in action and evaluate its impact on the system's behavior and overall QoS. This dashboard plays a vital role in offering insights into the system's adaptive mechanisms and their outcomes.

## 3 ARCHITECTURE AND DESIGN

SWITCH comprises core components like Managed System, Frontend, Environment Manager, and Managing System, each integral to the system's adaptability and user interaction. Unlike traditional SAS techniques that modify software architecture, SWITCH addresses specific ML challenges like handling data variability and ensuring model accuracy. This approach reflects a shift from architectural adjustments to dynamic ML model management, showcasing a novel aspect of adaptability in real-world MLS systems.

### 3.1 Managed System

*3.1.1* **Image Ingestion Service:** In real-world scenarios, especially for online-deployed Machine Learning Systems, user requests are processed asynchronously, i.e. continuously accepting them regardless of the processing time by model - a key feature of dynamic and responsive ML services. SWITCH emulates this real-world behavior in its *Image Ingestion Service*. This service, powered by FastAPI, Uvicorn[2] and python, receives image data from users (simulated by the *Load Simulator*) and stores it in the *Image Store* in a Base-64 Encoded format. This service efficiently handles concurrent, real-time data at variable rates, mirroring asynchronous user interaction for seamless subsequent processing.

*3.1.2* **Image Store:** *Image Store* in SWITCH functions as a dynamic queue within the local storage. It stores the incoming image data from the *Image Ingestion Service.* Images are queued here and are picked for processing based on their arrival order and then removed from the queue, implementing a first-in-first-out (FIFO) mechanism. At any given time, the number of images in the queue reflects pending images to be processed, providing a real-time view of the workload similar to operational queues in MLS deployments.

*3.1.3* **Data Preprocessor:** *Data Preprocessor* picks the oldest unprocessed image from the *Image Store* for preprocessing. It prepares the image data for object detection. It opens and loads image data from a byte array into memory for model processing, ensuring data readiness for model inference.

*3.1.4* **Model Loader:** *Model Loader* in SWITCH is a dynamic component that manages the *ML Model* in use as shown in Figure 1. It continuously monitors a specific file ('model.csv' in SWITCH) for indications of which model to load and process. This approach allows for real-time model switching based on external inputs, reflecting a key aspect of adaptability in real-world MLS systems. When SWITCH starts, it preloads all the models and stores them in the Model Repository as a dictionary, ready for switching. *Model Loader* ensures that the system is always ready to respond with the appropriate model as required.

*3.1.5* **Model Repository:** *Model Repository* of SWITCH houses YOLOv5nu, YOLOv5su, YOLOv5mu, YOLOv5lu & YOLOv5xu - all preloaded models of the YOLOv5u algorithm provided by Ultralytics [12], a state-of-the-art object detection system renowned for its accuracy and efficiency. YOLOv5u models, developed using the PyTorch framework and trained on the COCO dataset [16], are ready for deployment in the repository. The concept of 'preloading' models here means that each model is initialized and kept ready for immediate use, ensuring the system's adaptability and responsiveness to different object detection requirements. In addition to this, SWITCH's Model Repository allows users to integrate different types of object detection models.

*3.1.6* **ML Model**. In SWITCH, the *'ML Model'* refers to the currently active YOLOv5 model processing the image data. This model, selected by the *Model Loader*, is the primary driver of the object detection task within the system. It receives preprocessed images, applies the detection algorithms, and generates results, embodying the core functionality of an ML system in operation.

*3.1.7* **Post Processor:** *Post Processor* refines detection outcomes with a confidence score threshold (e.g., 0.35), focusing on desired classes (e.g., Humans, Cars). It computes total detections and average confidence. System metrics include the processing timestamp, count of processed requests (e.g., Request No. 370), current model name, model processing time, total time from image receipt to output (total_time), duration since project start (absolute time), and utility based on response time and confidence. System logs in JSON format are also generated for detailed performance insights.

*3.1.8* **Result Storage:** *Result Storage* is a temporary storage which manages processed data flow into the Elasticsearch-based *Data Store*, inside *Knowledge* in the managing system. REST API ensures seamless data transfer from the *Result Storage* to *Data Store*, critical for real-time performance understanding and adaptive decision-making. This integration enables SWITCH to maintain efficient storage, runtime observability and adaptability.

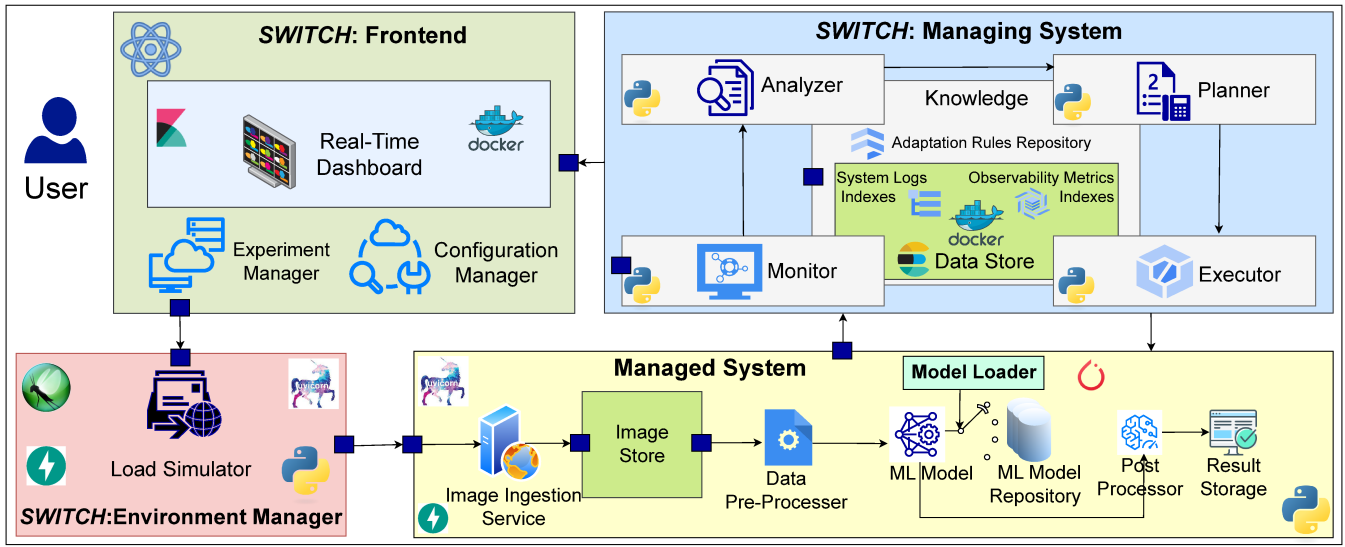

Figure 1: SWITCH : Architecture Diagram

## 3.2 Switch: Front-end

SWITCH's front-end, comprising the Experiment Manager, Configuration Manager, and Real-Time Dashboard, is the hub for user interaction, offering an intuitive and user-friendly experience.

*3.2.1 Experiment Manager & Configuration Manager:* Built with React, these components form the primary interface for user interaction. *Experiment Manager* facilitates the uploading of image data and interarrival rate files. Users can either upload .zip or can directly give the path of the image data folder stored locally. *Configuration Manager* is where users select or upload MAPE-K files for the managing system, determining the adaptation strategy. This includes choosing from predefined strategies like AdaMLs [15] or Naive [15], or uploading custom MAPE-K files. The backend, developed using FastAPI and hosted with Uvicorn as an ASGI server, manages interaction with the frontend UI through a series of API endpoints defined in FastAPI as shown in table 1. It enables operations such as starting and stopping processes, uploading data, and retrieving metrics, ensuring a seamless interaction between the user and the system.

*3.2.2 Real-Time Dashboard:* *Real-Time Dashboard*, developed using Kibana, is a key component of SWITCH's user interface. Integrated as an iframe within the React application and runs in a docker container, it provides real-time interactive visualizations of the latest system performance metrics, sourced directly from the Elasticsearch-based *Data Store* through REST API endpoints. Kibana's data visualization capabilities allow users to explore a wide array of metrics interactively. The *Real-Time Dashboard* enables users to analyze and learn about self-adaptation strategies in MLS through runtime model switching, thereby playing a crucial role in the adaptive process of the MLS. In conclusion, the front-end components of SWITCH —from the interactive React-based UI to the insightful Kibana *Real-Time Dashboard*—significantly enhance user engagement, providing critical insights into the system's performance and adaptive strategies in run-time.

## 3.3 SWITCH: Environment Manager

SWITCH's Environment Manager features a *Load Simulator*, essential for replicating real-world API traffic showing load uncertainty in the environment and testing the system's adaptability. It receives image data and interarrival rate files from the *Experiment Manager* of the SWITCH frontend through FastAPI endpoints. This component uses Locust, an open-source load testing tool, to emulate user behavior and manage incoming image data distribution. Locust's Python script simulates user interactions, which is crucial for evaluating SWITCH's performance under various operational loads. The simulator leverages interarrival rate data from user inputs, like the FIFA98 World Cup logs [22], to create realistic traffic patterns. These patterns, characterized by time gaps between image uploads, test the system's responsiveness to fluctuating and peak load conditions. These scripts direct traffic to *Image Ingestion Service* through FastAPI within the Managed System, ensuring realistic and varied testing scenarios. By mirroring real-world user behaviors and traffic scenarios, this setup is essential for practitioners to analyze and enhance MLS systems' adaptive capabilities by assessing the impact of their adaptation strategies in practical environments.

## 3.4 SWITCH: Managing System

*3.4.1 Knowledge:* *Knowledge* component within Switch's Managing System plays a central role in adaptive decision-making of model switching. It contains *Adaptation Rules Repository*, storing adaptation rules by user for MAPE loop. Its Elasticsearch-based *Data Store*, which is adept at handling large-scale data processing and analytics, runs in a docker container. It continually receives data from the Managed System via REST API and stores in two indexes: - *new_logs* for troubleshooting and performance analysis in *System Logs Indexes* as JSON documents. - *final_metrics* for monitoring system performance and guiding adaptations in - *Observability Metrics Indexes* as JSON documents. Elasticsearch's JSON-based REST API facilitates CRUD operations and data searches, enhancing the system's adaptive capabilities. *Knowledge* provides these metrics to *Real-Time Dashboard* for visualization and to *Monitor* for monitoring the system's realtime performance through REST API.

*3.4.2 Self-Adaptation Through MAPE-K Framework:* Switch's self-adaptive capabilities are primarily demonstrated using the MAPE-K framework, which is a standard approach for implementing self-adaptive systems: i) **Monitor** retrieves metrics from **Knowledge** using API calls for real-time system monitoring; ii) **Analyzer** analyzes the monitored data to assess whether adaptation is necessary; iii) **Planner** develops strategies based on the analysis for potential adaptations; iv) **Executor** executes the adaptation strategies, influencing the managed system as needed. SWITCH offers flexibility by allowing users to use, customize, or create adaptation strategies within the MAPE-K framework. This versatility makes it an ideal tool for exploring various model-switching approaches in self-adaptation for MLS. SWITCH's API, facilitates interaction with the system for custom strategy implementation. Table 1 lists the key API endpoints that users can leverage to programmatically interact with SWITCH. These endpoints allow real-time adaptation and monitoring.

## 4 SYSTEM USAGE & ADAPTATION

### 4.1 System Usage:

SWITCH, an exemplar for Machine Learning-Enabled Systems (MLS), offers an intuitive and straightforward user experience. It reflects real-world scenarios of MLS in its design and operation. Upon initiating SWITCH, the user's actions set in motion a series of automated processes. Docker Compose uses containers to launch Elasticsearch and Kibana services to establish the backend for data visualization and storage. Concurrently, the backend services of SWITCH are brought online through the execution of the Node.py script, creating essential links between the system's front and back ends. The React application, comprising the Experiment Manager and Configuration Manager, becomes operational and presents the user with SWITCH's home page, an interactive web interface as shown in figure 2 and preloads all ML models in the repository through *Model Loader*. Once experiments commence, the integrated Kibana dashboard within the React application becomes accessible, offering real-time insights into system performance.

In SWITCH, users engage in the following activities on the home page to start the experiment: **i) Upload Image Data:** Users can upload images either as a .zip file or directly from a local folder.

| API Endpoint | Description |
|---|---|
| /api/stopProcess | Stops the current process |
| /api/downloadData | Downloads logs and metrics |
| /api/latest_metrics_data | Retrieves the latest metrics |
| /api/latest_logs | Retrieves the latest system logs |
| /api/changeKnowledge | Changes adaptation knowledge |
| /api/upload | Uploads input to the server |

Table 1: API endpoints and their descriptions.

Figure 2: SWITCH User Interface: Home Page

This versatility enables SWITCH to handle data directly from the user's environment; **ii) Inter-arrival Rate File:** Users upload a .csv file for inter-arrival rates, allowing them to test the system's performance under varied real-world conditions; **iii) Experiment ID:** Users assign an ID to their experiment, under which all related logs and metrics are organized and stored; **iv) Select Self-Adaptation Strategy:** Through a drop-down menu, users can choose from a range of self-adaptation strategies detailed in the subsequent subsection, tailoring the system's adaptation approach to their needs.

### 4.2 Adaptation Strategies

The switch incorporates a range of self-adaptation strategies, notably the NAIVE and AdaMLS approaches, each uniquely enhancing the system's adaptability as explained in [15].

**Single Model Strategies:** These strategies (no switching) involve running a single YOLOv5u model variant throughout the experiment. Users can choose from five YOLOv5u models, each catering to specific performance requirements.

**NAIVE and Modified NAIVE:** The NAIVE approach, based on the incoming rate of images, switches between models to balance speed and accuracy. For instance, it uses YOLOv5nu (nano) for higher rates (15-30 images/sec) and YOLOv5xu for lower rates (below 2 images/sec). The Modified NAIVE approach allows users to customize these threshold values and adapt the strategy to their specific needs.

**AdaMLS:** This novel approach, based on unsupervised learning, assesses the capabilities of different models in real-time and selects the one offering the highest confidence score while meeting the target response time. AdaMLS's detailed methodology and its impact on enhancing Quality of Service are discussed in [15]. SWITCH provides AdaMLS implementation, enabling users to experiment with this advanced adaptation strategy.

**Custom MAPE-K Strategies:** SWITCH is designed to be flexible, allowing users to develop and deploy their own MAPE-K strategies with ease. The system's user-friendly interface and accessible APIs make monitoring and executing custom strategies straightforward. Below is an example showing how users can set up both monitoring and model-switching functionalities.

**Listing 1: Custom MAPE-K Strategy Implementation**

```
# Monitoring Metrics from Elasticsearch
def fetch_metrics(index_name, fields, num_docs_to_fetch):
    query = {"size": num_docs_to_fetch, "sort": [{"log_id": {"order": "desc"}}]}
    response = es.search(index=index_name, body=query)
    # Example field processing for 'model_processing_time'
    processed_metrics = process_response(response, fields)
    return processed_metrics

def process_response(response, fields):
    # Logic to process fields like 'model_processing_time' from response
    return averaged_metrics

# Switching Model based on a condition
def switch_model(model_name):
    with open("model.csv", "w") as file:
        file.write(model_name)
    # SWITCH checks 'model.csv' and updates the model accordingly
```

In this implementation, the 'fetch_metrics' function retrieves desired metrics from Elasticsearch, and the 'switch_model' function updates the 'model.csv' file to switch models. SWITCH continuously monitors this file and adapt the active model as specified, demonstrating its capability for real-time, dynamic self-adaptation.

## 5 EMPIRICAL EVALUATION

Switch employs YOLOv5u models [12] for a wide range of object detection scenarios. These models are renowned for their accuracy and efficiency and are trained on the COCO dataset [16], which includes 80 object categories.

**Case Study:** *1. General Detection:* Utilizing 10,000 images from the COCO 2017 Unlabelled dataset (1.6 GB) [16], Switch handles a wide variety of 80 categories, showcasing its capacity to deal with diverse general object detection tasks. *Note:* Evaluations for *2. Crowd Detection* and *3. Traffic Detection* was also conducted, focusing on different scenarios and datasets. Results and analyses for these cases are available on our GitHub repository[5].

**Customizing Object Detection:** SWITCH enables easy customization for detection tasks. Users can modify the detection process by altering the process.py file. For instance, filtering results based on a confidence threshold (e.g., 0.35) and desired class IDs allows targeted detection for classes like 'crowd' or 'vehicle'.*Example: For 'crowd' class: if confidences[i] >= 0.35 && class_list[i] == 0 {...}*

**System Requirements:** SWITCH can be deployed on any laptop or PC capable of running Docker and supporting Linux. For detailed technical requirements and setup instructions, please refer to our GitHub repository.

### 5.1 Evaluation using AdaMLS Approach

The AdaMLS approach from [15] was directly applied in SWITCH, with tests on a 12th Gen Intel(R) Core(TM) i5-12500H -12 CORE system. We primarily focused on General Object Detection, using load conditions simulated with scaled FIFA98 logs [22]. AdaMLS's rules, which can be recalibrated based on different setups, were applied in SWITCH. For a comprehensive interpretation of the AdaMLS results, refer to the AdaMLS paper [15].

[5]https://github.com/sa4s-serc/switch

**Comparison of Model Switching Approaches:** To illustrate the effectiveness of model switching strategies in SWITCH, we conducted a comparison between the AdaMLS and a baseline 'Nano Model' approach. The latter represents a non-switching scenario for contrast. The following table 2 presents the comparison in terms of various performance metrics for General Object Detection:

**Table 2: Comparison of General Object Detection using AdaMLS Approach and Nano Model-(No Switching)**

| Metric | AdaMLS | Nano Model |
|---|---|---|
| Total Images Processed | 10000 | 10000 |
| Average Confidence Score | 0.7 | 0.65 |
| Average CPU Consumption | 20 | 20.14 |
| Total Objects Detected | 47026 | 37829 |
| Average Model Processing Time (s) | 0.033 | 0.015 |
| Average Image Processing Time (s) | 0.25 | 0.35 |

The results from the AdaMLS application, as shown in the Table 2 and Figure 3, illustrate the dynamic model switching capabilities of SWITCH. The table indicates that in 10,000 processed images, each containing multiple detectable objects, a total of 47,026 objects were detected. 'Average Model Processing Time' refers to the duration for which an image is processed within the model itself, while 'Average Image Processing Time' represents the end-to-end life-cycle of an image — from queueing to processing completion.

Figure 3 provides insights into the varying request rates over time and how AdaMLS smartly switches between models in response. When the load is high, SWITCH quickly transitions to the Nano Model for faster processing. Conversely, during periods of lower request rates, it opts for models with higher accuracy, achieving better confidence scores. In this way, SWITCH is used to test approaches. This adaptability demonstrates the critical need for smart model switching strategies in MLS, as it allows for a balance between speed and accuracy based on real-time demands.

**Effectiveness of SWITCH:** SWITCH preloads 5 YOLOv5u models, consuming 21.3 to 34.8 CPU, with load time of around 0.25 seconds and energy usage of about 10.56 joules. Model switching is executed in approximately 100 microseconds on average.

**Real-time Dashboard:** An excerpt from SWITCH's dashboard, depicted in Figure 4, showcases real-time metrics processing time versus requests processed. It also offers key performance indicators and various dynamic visual elements like pie charts and bar plots for a clear, real-time overview. This user-friendly dashboard enables users to effectively analyze and adjust their custom strategies for machine learning systems, supporting the development and refinement of new adaptive approaches in MLS.

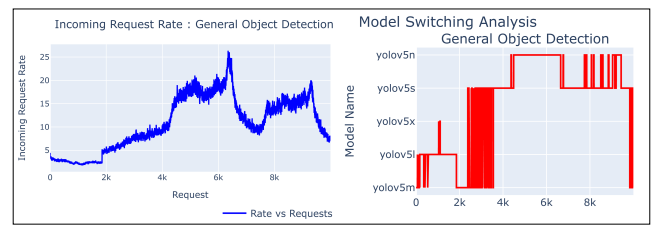

**Figure 3: SWITCH : Request Rate and Model Switching**

## 5.2 Technical Challenges & Solutions

In SWITCH, we've preloaded various YOLOv5u models to enable quick switching based on real-time performance, addressing the challenge of working with different ML models. For efficient handling of large volumes of image data, images are stored locally for swift processing, and Elasticsearch is used for dashboard updates, ensuring smooth data management. The dashboard, developed with Kibana and Elasticsearch, offers an intuitive and interactive experience, making real-time data easily understandable for users. Although SWITCH is designed for Linux environments, it can also be used on non-Linux systems through a Linux virtual environment, enhancing its compatibility across different computer types.

By tackling these challenges, we've made SWITCH a more flexible tool. It's now better suited for research and practical applications in the field of MLS that adapt themselves based on changing conditions.

## 6 RELATED WORK

Self-adaptive systems (SAS) traditionally focused on managing systems without embedded ML models [24]. These systems utilized tactics like architectural changes [8] or service quality adjustments [6] [19], in response to environmental changes [14]. However, the integration of ML components into managed systems introduces new challenges, especially regarding the variability of properties such as accuracy [4] [5] and quality of service [21] [3] [17]. The concept of self-adaptation in ML was initially introduced in [1], [20]. Further discussions and explorations in this direction are highlighted in [4] [5]. The field has seen approaches like AdaMLS improve Quality of Service (QoS) in ML-enabled systems through adaptive strategies [15].

Additionally, recent advancements in object detection, primarily focused on individual machine learning model enhancements [7, 9–11, 13, 25], have overlooked broader system adaptability aspects. Despite these developments, a review of the SEAMS community's repository of exemplars i.e. self-adaptive.org reveals a significant gap in tools specifically for ML-enabled systems. SWITCH fills this gap, offering a platform for dynamic model switching and enhanced user experience, distinguishing itself from existing tools like SWIM [18]. This makes SWITCH a pioneering exemplar in SAS having a self adaptive ML-enabled system, enabling research, exploration & experimentation with self-adaptation strategies in a real-world ML context.

## 7 CONCLUSION & FUTURE WORK

This paper presents SWITCH, a pioneering tool designed for MLS, with a current focus on runtime model-switching as its core functionality. SWITCH stands out for its principles of self-adaptation and dynamic responsiveness, featuring a scalable architecture, a user-friendly interface, and a versatile dashboard that facilitates both experimentation and practical insights into MLS. Through empirical evaluations in object detection scenarios, SWITCH has demonstrated its adaptability and performance, showcasing the efficient management of varying conditions in MLS, underlines its potential and sets the stage for future enhancements.

In addition to its technical contributions, SWITCH serves as a valuable educational and research tool. It provides a platform for researchers, practitioners, and students to explore and understand the intricacies of self-adaptation in MLS. By enabling hands-on experimentation with different scenarios and strategies, SWITCH inspires innovative solutions and approaches in the evolving landscape of self-adaptive systems with a deeper understanding of MLS.

While currently focused on model switching, future iterations of SWITCH aim to explore a broader spectrum of MLS tasks including retraining, transfer learning and user-driven customization beyond the MAPE-K framework. Additionally, further empirical studies and adaptations to address emerging MLS challenges and uncertainties are anticipated. Ultimately, the insights gained from SWITCH will guide the development of more versatile systems, exploring various adaptation strategies in MLS to meet both present and future demands in the field of self-adaptive systems and machine learning. This positions SWITCH as a significant contribution to the domain, emphasizing the critical role of adaptability in advancing MLS.

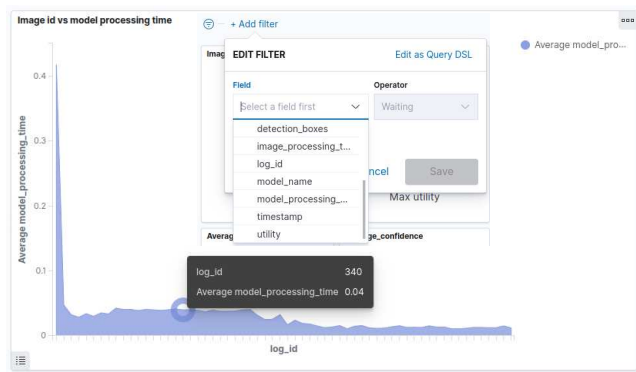

Figure 4: SWITCH : Runtime Dashboard Excerpt